\documentclass[prl,aps,twocolumn,superscriptaddress]{revtex4-1}

\usepackage{color,amsmath,amssymb,amsfonts,graphicx,tabularx}
\usepackage{ulem}
\usepackage[unicode=true,colorlinks=true]{hyperref}

\hypersetup{linkcolor=blue,citecolor=blue,urlcolor=blue}

\begin{document}

\title{SU(3) trimer resonating-valence-bond state on the square lattice}

\author{Xiao-Yu Dong}
\affiliation{Max-Planck-Institut f{\"u}r Physik komplexer Systeme, N{\"o}thnitzer Stra{\ss}e 38, 01187 Dresden, Germany}
\affiliation{Department of Physics and Astronomy, California State University, Northridge, California 91330, USA}

\author{Ji-Yao Chen}
\email{Ji-Yao.Chen@irsamc.ups-tlse.fr}
\affiliation{Laboratoire de Physique Th{\'e}orique, C. N. R. S. and Universit\'e de Toulouse, 31062 Toulouse, France}

\author{Hong-Hao Tu}
\affiliation{Institut f\"ur Theoretische Physik, Technische Universit\"at Dresden, 01062 Dresden, Germany}

\begin{abstract}
We propose and study an SU(3) trimer resonating-valence-bond (tRVB) state with $C_{4v}$ point-group symmetry on the square lattice. By devising a projected entangled-pair state representation, we show that all (connected) correlation functions between local operators in this SU(3) tRVB state decay exponentially, indicating its gapped nature. We further calculate the modular $S$ and $T$ matrices by constructing all nine topological sectors on a torus and establish the existence of $\mathbb{Z}_3$ topological order in this SU(3) tRVB state.
\end{abstract}

\maketitle

\date{\today}

\textit{Introduction} --- The search for topological phases of matter in realistic systems has attracted tremendous interest in recent years~\cite{wen2004}. The celebrated examples that have been realized in experiments are the integer and fractional quantum Hall states~\cite{klitzing1980,tsui1982,laughlin1983}. These exotic phases are not amenable to Landau's local order parameter description and are instead characterized by the topological characters, such as protected ground-state degeneracy on a torus, energy gap above the ground-state manifold, gapless edge excitations, and quasiparticles with fractionalized quantum numbers and anyonic statistics. The robustness to external perturbations renders topological states potential applications in quantum information processing~\cite{kitaev2003,freedman2002,nayak2008}.

Frustrated magnetic materials form another important platform for studying topological phases. Much effort in this area has been devoted to finding spin liquids that do not order down to zero temperature~\cite{zhou2017}. From the theoretical viewpoint, a large fraction of spin liquids, when a gap is present, are topological phases enriched by spin rotational and/or lattice symmetries. Further classification of these spin liquids can be carried out based on the type of topological order and present symmetries. However, for a given spin model, it is usually notoriously difficult to determine whether the ground state is a spin liquid, let alone how it fits into the classification scheme. In this regard, wave functions capturing the essential physics provide an important link between the classification and the microscopic models. One such example is the spin-1/2 nearest-neighbor resonating-valence-bond (RVB) state on the triangular lattice~\cite{moessner2001}, which is a $\mathbb{Z}_2$ spin liquid realizing the simplest $\mathbb{Z}_2$ topological order. The Kalmeyer-Laughlin state is another example which is a chiral spin liquid with broken time-reversal symmetry~\cite{kalmeyer1987}.

In recent years, rapid development in the experimental manipulation of cold atoms with SU($N$)-symmetric exchange interactions has attracted considerable interest~\cite{zhang2014,scazza2014}, partly because of the potential of hosting chiral spin liquids~\cite{hermele2009}. Further theoretical and numerical investigations on SU($N$) quantum magnets have found a number of magnetically ordered and topological phases~\cite{bauer2012,corboz2013,wu2016,pimenov2017} which do not seem to have obvious SU(2) counterparts. For studying possible novel phases in SU($N$) magnets, it is thus expected that the wave function approach can provide useful information complementing field theoretical and numerical approaches. However, except for the SU($N$) generalization of the Kalmeyer-Laughlin state~\cite{tu2014,bondesan2014}, wave functions for SU($N$) magnets remain largely unexplored so far.

In this work, we propose and study an SU(3) trimer resonating-valence-bond (tRVB) state with $C_{4v}$ point-group symmetry on the square lattice. The tRVB state, with its elementary building block being 90-degree bent SU(3) trimer singlets extending over three adjacent sites, is an equal-weight superposition of trimer coverings on the lattice. Unlike the SU(2) spin-1/2 nearest-neighbor RVB state which is a gapless spin liquid on the square lattice~\cite{albuquerque2010, tang2011}, we show that the SU(3) tRVB state is a gapped spin liquid with $\mathbb{Z}_3$ topological order. For that we devise a projected entangled-pair state (PEPS) representation, and characterize the state with powerful PEPS techniques. We show that (i) all (connected) correlation functions between local operators in the tRVB state decay exponentially; (ii) the calculated modular $S$ and $T$ matrices are in agreement with the $\mathbb{Z}_3$ topological order. These results demonstrate that the SU(3) tRVB state on the square lattice is indeed a $\mathbb{Z}_3$ spin liquid.

\begin{figure}
\includegraphics[width=0.49\textwidth]{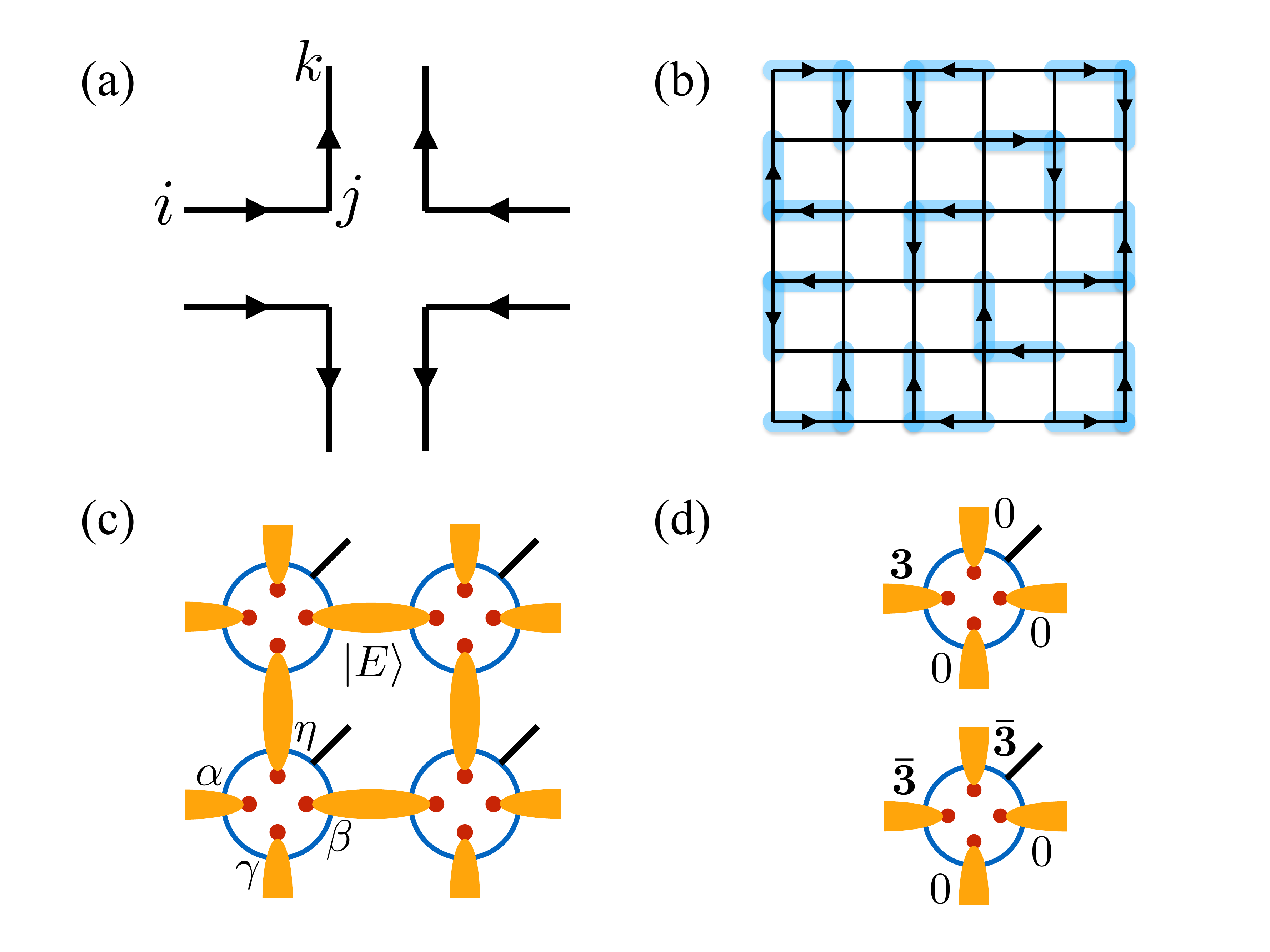}
\caption{(a) Schematics of the SU(3) trimers. (b) Typical trimer covering configuration on the square lattice. (c) Projected entangled-pair state representation of the SU(3) tRVB state. (d) Two kinds of projectors mapping virtual states to physical states.}
\label{Figure1}
\end{figure}

\textit{SU(3) tRVB state} --- Let us consider a square lattice with spins on each site transforming under the fundamental representation (denoted by $\mathbf{3}$) of SU(3). The local spin basis is defined by $|a\rangle$, where $a=1,2,3$. The elementary building block of the SU(3) tRVB state is an SU(3) trimer singlet formed among three sites: $|\mathrm{trimer}\rangle_{ijk} = \sum_{a,b,c\in \{1,2,3\}}\varepsilon_{abc}|a\rangle_i|b\rangle_j|c\rangle_k$ , where $i,j,k$ stand for lattice sites and $\varepsilon_{abc}$ is a totally antisymmetric tensor with $\varepsilon_{123}=1$. For such trimer, it is convenient to assign an orientation as $i\rightarrow j\rightarrow k$. For our purpose, we only consider four kinds of ``short-range'' bent trimers for which both $(i,j)$ and $(j,k)$ are nearest neighbors and the angle between two orientations $i \rightarrow j$ and $j\rightarrow k$ is 90 degrees [see Fig.~\ref{Figure1}(a)].

The SU(3) tRVB state of our interest is an equal weight superposition of trimer coverings on the square lattice; see Fig.~\ref{Figure1}(b) for an example of the trimer covering configuration. The relative sign of trimer covering configurations is fixed by the local orientations of the trimers, i.e., only trimers with orientations shown in Fig.~\ref{Figure1}(a) are allowed. It is then straightforward to verify that the tRVB state so obtained respects the full $C_{4v}$ symmetry of the square lattice. Similar to the non-orthogonality of SU(2) valence-bond dimer coverings, the SU(3) trimer covering configurations do not form an orthogonal basis either. Thus, our tRVB is different from a recent proposal \cite{lee2017} of tRVB wave function which consists of orthogonal trimer configurations, in the same sense as the difference between the spin-1/2 nearest-neighbor RVB state \cite{kivelson1987} and the Rokhsar-Kivelson wave function \cite{rokhsar1988}. Due to the non-orthogonality, there is \textit{a priori} no reason to suppose that the correlations of the SU(3) tRVB state are similar to its classical analog~\cite{froboese1996}.

\textit{PEPS representation} --- In order to characterize the SU(3) tRVB state, we switch to its PEPS representation. Similar strategy has been proven very successful in characterizing the spin-1/2 RVB~\cite{verstraete2006,poilblanc2012,schuch2012,wang2013,chen2018} and spin-1 resonating AKLT-loop states~\cite{yao2010,li2014} on various lattices. Following the projective construction of PEPS, we introduce at every site four virtual particles, each of which supports a 7-dimensional auxiliary Hilbert space $\mathcal{V}_A$ with basis vectors $|0\rangle$ belonging to the SU(3) trivial representation, and $\{|1\rangle, |2\rangle, |3\rangle \}$ ($\{|\bar{1}\rangle, |\bar{2}\rangle,|\bar{3}\rangle\}$) transforming under the fundamental (anti-fundamental) representation $\mathbf{3}$ ($\bar{\mathbf{3}}$), respectively. Each pair of virtual particles between adjacent sites forms a maximally entangled state [see Fig.~\ref{Figure1}(c)]
\begin{equation}
|E\rangle = |00\rangle+|1\bar{1}\rangle + |2\bar{2}\rangle + |3\bar{3}\rangle + |\bar{1}1\rangle + |\bar{2}2\rangle + |\bar{3}3\rangle.
\label{eq:bond1}
\end{equation}
For later purpose we compactly write the maximally entangled state (\ref{eq:bond1}) between sites $i_1$ and $i_2$ as
\begin{equation}
|E\rangle_{i_1,i_2} = \sum_{\alpha,\beta\in \mathcal{V}_A} E_{\alpha,\beta} |\alpha\rangle_{i_1}|\beta\rangle_{i_2},
\label{eq:bond2}
\end{equation}
where the nonvanishing entries of $E_{\alpha,\beta}$ can be obtained from (\ref{eq:bond1}).

To recover the physical Hilbert space, the four virtual states at each site are projected back to the physical state by a projector $\hat{P}$, defined by
\begin{equation}
\hat{P} = \sum_{a\in \mathcal{V}}\sum_{\alpha,\beta,\eta,\gamma\in \mathcal{V}_A} \mathcal{P}^a_{\alpha,\beta,\eta,\gamma} |a\rangle\langle \alpha,\beta,\eta,\gamma|,
\label{eq:projector1}
\end{equation}
where $\alpha,\beta,\eta,\gamma$ are assigned for the virtual states at left, right, up, down positions [see Fig.~\ref{Figure1}(c)], $\mathcal{V}$ is the physical Hilbert space on each site, and $\mathcal{P}^a_{\alpha,\beta,\eta,\gamma}$ is a tensor to be specified below.
To reproduce the tRVB state, we decompose the projector $\hat{P}$ into two parts,
\begin{equation}
\hat{P}=\hat{P}_1+\hat{P}_2,
\label{eq:projector2}
\end{equation}
where $\hat{P}_1$ identifies one of the virtual state in $\mathbf{3}$ as the physical state [the rest three virtual particles are in the trivial representaiton; see upper panel in Fig.~\ref{Figure1}(d) for an example],
\begin{eqnarray}
\hat{P}_1 &=& \sum_{a\in \mathcal{V}}\sum_{\alpha,\beta,\eta,\gamma\in \mathcal{V}_A} [(\delta_{\alpha,a}\delta_{\beta,0} + \delta_{\alpha,0}\delta_{\beta,a})\delta_{\eta,0}\delta_{\gamma,0} \nonumber\\
&& -\delta_{\alpha,0}\delta_{\beta,0}(\delta_{\eta,a}\delta_{\gamma,0}+\delta_{\eta,0}\delta_{\gamma,a})]
|a\rangle\langle \alpha,\beta,\eta,\gamma|,
\label{eq:projector3}
\end{eqnarray}
and $\hat{P}_2$ maps two adjacent virtual states in $\bar{\mathbf{3}}$ into the physical state [the rest two virtual particles are in the trivial representation; see lower panel in Fig.~\ref{Figure1}(d)],
\begin{eqnarray}
\hat{P}_2 &=& \sum_{a\in \mathcal{V}}\sum_{\alpha,\beta,\eta,\gamma\in \mathcal{V}_A} \sum_{\bar{M},\bar{N}\in (\bar{1},\bar{2},\bar{3})}\varepsilon_{a M N}
(\delta_{\alpha,\bar{M}}\delta_{\beta,0}\delta_{\eta,\bar{N}}\delta_{\gamma,0}  \nonumber\\
&& \quad +\delta_{\alpha,\bar{M}}\delta_{\beta,0}\delta_{\eta,0}\delta_{\gamma,\bar{N}}
+ \delta_{\alpha,0}\delta_{\beta,\bar{M}}\delta_{\eta,\bar{N}}\delta_{\gamma,0}   \nonumber\\
&& \quad + \delta_{\alpha,0}\delta_{\beta,\bar{M}}\delta_{\eta,0}\delta_{\gamma,\bar{N}})
|a\rangle\langle \alpha,\beta,\eta,\gamma|.
\label{eq:projector4}
\end{eqnarray}
The tensor $\mathcal{P}^a_{\alpha,\beta,\eta,\gamma}$ in (\ref{eq:projector1}) is thus defined through the sum of tensor entries in (\ref{eq:projector3}) and (\ref{eq:projector4}). It is also easy to verify that both $\hat{P}_{1}$ and $\hat{P}_{2}$ belong to the $B_1$ irreducible representation of the $C_{4v}$ point-group symmetry~\cite{landau1977}. Here we would like to mention that, the linear trimer configuration where the three neighboring sites forming singlet are on a straight line is excluded since it does not belong to the $B_1$ irreducible representation of $C_{4v}$.

With the PEPS projector and the virtual bonds in hand, the PEPS for the SU(3) tRVB state is obtained by applying the product of projectors to the virtual bonds
\begin{equation}
|\psi\rangle = \bigotimes_{i=1}^{N}\hat{P}^{(i)}\bigotimes_{\langle i_1,i_2\rangle}|E\rangle_{i_1,i_2},
\label{eq:peps}
\end{equation}
which is shown in Fig.~\ref{Figure1}(c).
By construction, each trimer consists of three sites with $\hat{P}_2$ acting on the middle site and $\hat{P}_1$ acting on the two end sites. Only the configurations of trimer coverings, in which each site belongs to one and only one trimer, have non-zero weight in $|\psi\rangle$, and all trimer covering configurations have equal weights with sign conventions consistent with the definition in Fig.~\ref{Figure1}(a). The state will not change if there is a nonzero real coefficient $\lambda$ in $\hat{P} = \hat{P}_1 + \lambda  \hat{P}_2$.

An alternative representation of $|\psi\rangle$ that will also be used below is obtained by eliminating the virtual states in (\ref{eq:peps}), and the wave-function amplitude takes the form of a tensor network, $\psi(\ldots,a_i,\ldots) = \sum_{\{\ldots \alpha_i,\beta_i,\eta_i,\gamma_i \ldots\}} (\cdots \mathcal{A}^{a_i}_{\alpha_i,\beta_i,\eta_i,\gamma_i} \cdots)$, where $\mathcal{A}^a_{\alpha,\beta,\eta,\gamma}$ is given in terms of the virtual bond (\ref{eq:bond2}) and the PEPS projector (\ref{eq:projector1}) as
\begin{equation}
\mathcal{A}^a_{\alpha,\beta,\eta,\gamma} = \sum_{\beta',\eta'\in \mathcal{V}_A}\mathcal{P}^a_{\alpha,\beta',\eta',\gamma}E_{\beta',\beta}E_{\eta',\eta}.
\label{eq:Atensor}
\end{equation}

\begin{figure}
\includegraphics[width=0.49\textwidth]{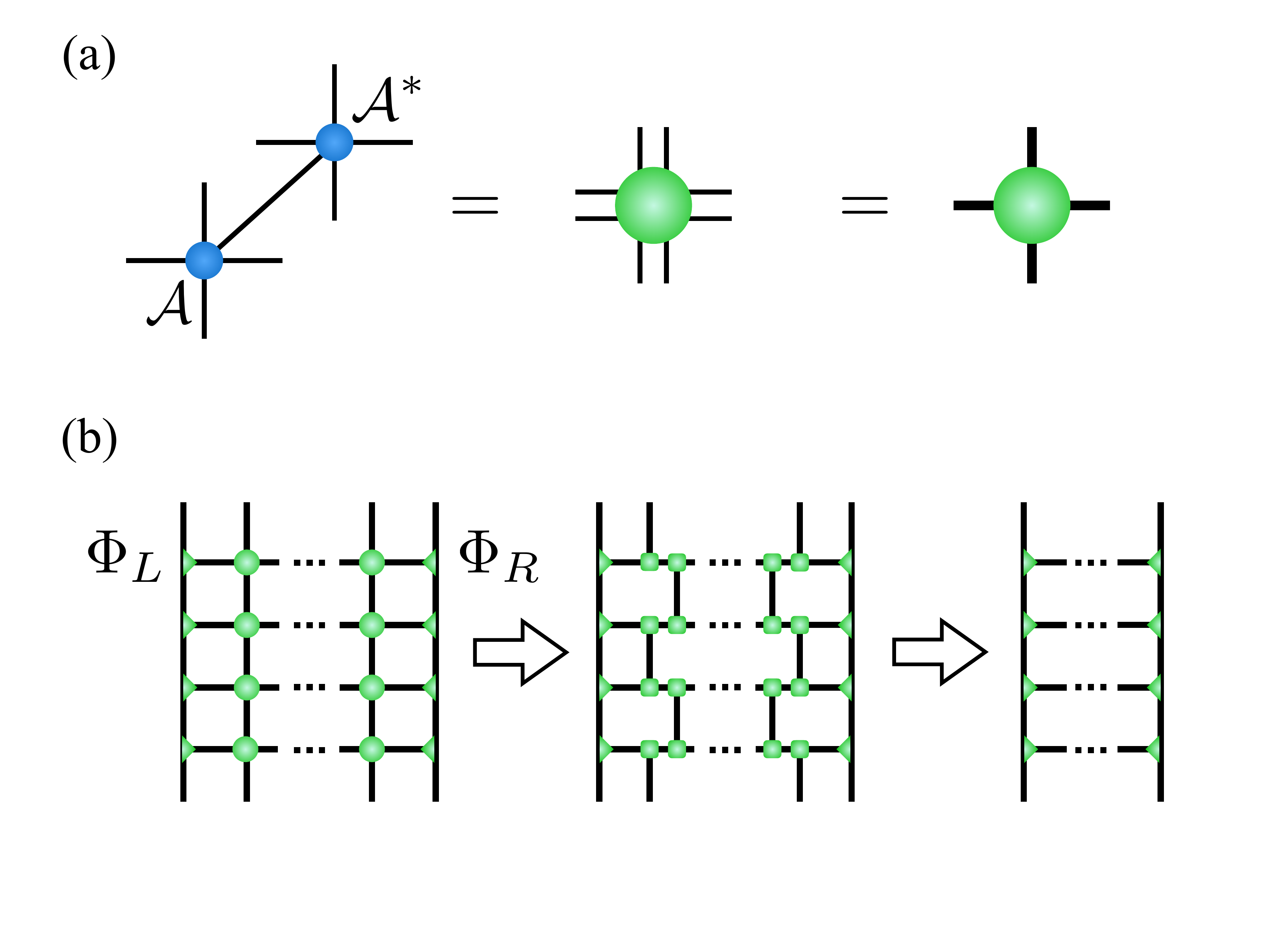}
\caption{(a) Each of the local tenor $\mathcal{A}$ of iPEPS is contracted with its complex conjugate by the physical index to form the bi-layer tensor network. (b) Schematic plot of the contraction of bi-layer tensor network based on boundary iMPS.}
\label{FigureS1}
\end{figure}

\begin{figure}
\includegraphics[width=0.49\textwidth]{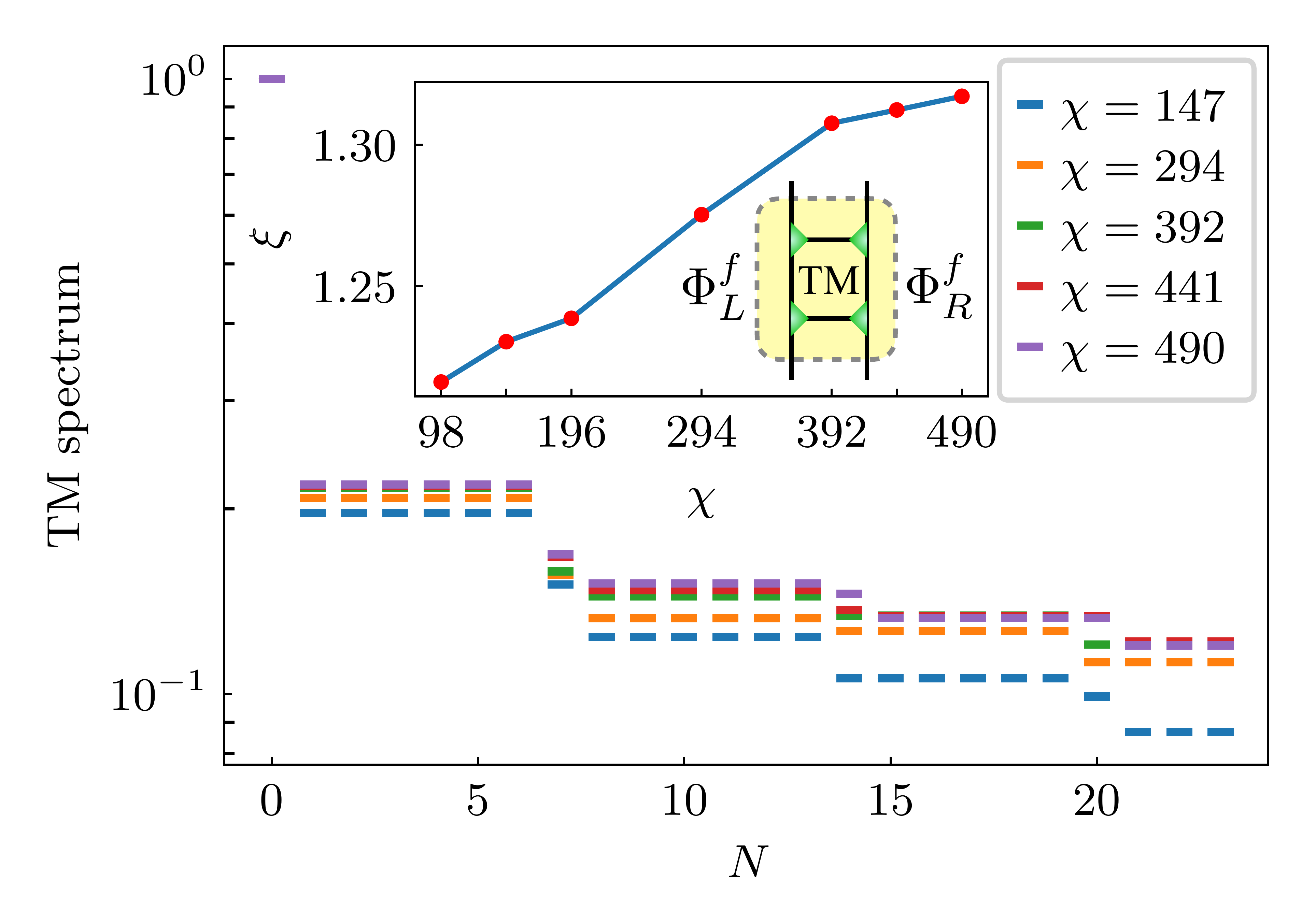}
\caption{The spectrum of the transfer matrix (only the absolute values are shown). The eigenvalues are normalized such that the largest eigenvalue is $1$. The inset defines the transfer matrix for a two-site unit cell and shows the upper bound of the correlation length $\xi$ versus the truncation bond dimension $\chi$.}
\label{Figure2}
\end{figure}

\textit{Correlation functions} --- The PEPS formulation is particularly convenient for calculating correlation functions characterizing the tRVB state.
The norm of the PEPS is a bi-layer tensor network obtained by contracting the physical indices of the PEPS (ket layer) and its complex conjugate (bra layer). The expectation value of local operators or their correlators is a similar bi-layer tensor network with local operators being inserted between the ket and bra layers. Here we consider an infinite system and contract the bi-layer tensor networks by using boundary infinite matrix-product state (iMPS) method~\cite{jordan2008}: the left (right) environment of the bi-layer tensor networks is represented by a boundary iMPS $\Phi_\mathrm{L}$ ($\Phi_\mathrm{R}$). One column of tensors, formed by a bi-layer element shown in Fig.~\ref{FigureS1}(a), defines an infinite matrix-product operator (iMPO). The environments, represented by two iMPSs, are obtained by iteratively applying the iMPO to initial boundary iMPSs. After a few iteration steps, it is necessary to truncate the bond dimension of the iMPSs, the maximum of which is denoted by the truncation bond dimension $\chi$. As shown in Fig.~\ref{FigureS1}(b), the iMPSs are obtained by using the infinite time-evolving block decimation algorithm (iTEBD) ~\cite{vidal2007,orus2009}, which determines the iMPSs as fixed points of the iteration.
Once converged iMPSs $\Phi_\mathrm{L}^f$ and $\Phi_\mathrm{R}^f$ ($f$ stands for ``fixed point'') are obtained, a transfer matrix (TM) can be constructed as shown in the inset of Fig.~\ref{Figure2} (note that the iTEBD algorithm produces an enlarged two-site unit cell, and we have checked that the one-site translational symmetry of physical observables is not broken with large enough $\chi$). For the tRVB state, we find that the largest eigenvalue of the TM is unique (normalized to unity) and there is a finite gap between the largest eigenvalue and the second largest eigenvalue (see Fig.~\ref{Figure2} for the saturation of the gap when increasing $\chi$ up to $\chi=490$ with the largest truncation error $\sim4\times 10^{-6}$). Thus, there is no spontaneous symmetry breaking in the tRVB state and all (connected) correlation functions decay exponentially with a correlation length upper bounded by $\xi=-2/\log(|\lambda|) \sim 1.317$ (estimate with $\chi=490$), where the factor of $2$ appears due to the two-site unit cell in the fixed-point iMPSs, and $\lambda$ is the second largest eigenvalue of the TM. Because of the $C_{4v}$ symmetry of the tRVB state, the correlation lengths are the same along horizontal and vertical directions.

\begin{figure}
\includegraphics[width=0.49\textwidth]{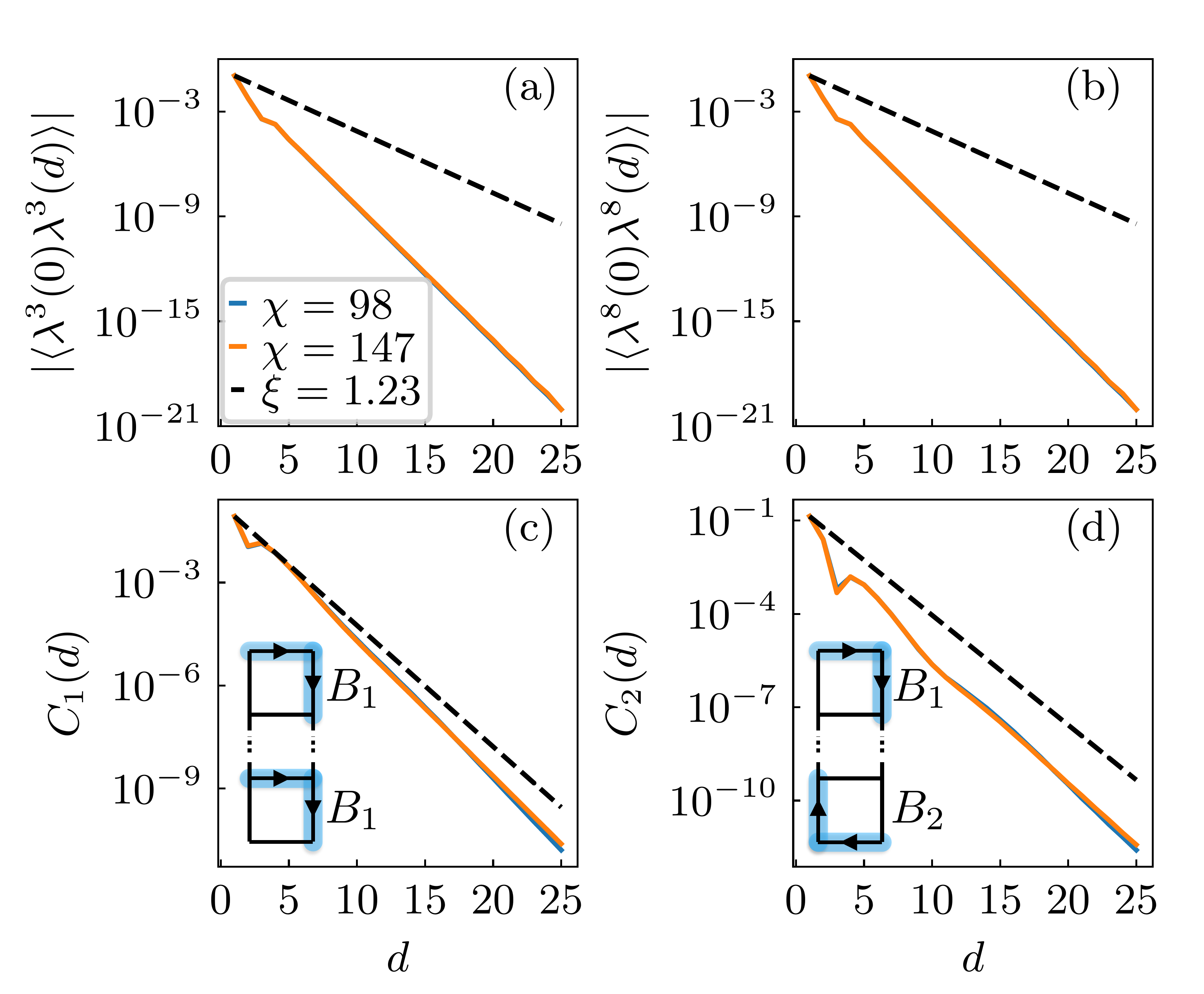}
\caption{The correlation functions (a) $|\langle \lambda^3(0)\lambda^3(d)\rangle|$, (b) $|\langle \lambda^8(0)\lambda^8(d)\rangle|$, (c) $C_1(d)=|\langle B_1(0)B_1(d)\rangle-\langle B_1(0)\rangle\langle B_1(d)\rangle|$, and (d) $C_2(d)=|\langle B_1(0)B_2(d)\rangle-\langle B_1(0)\rangle\langle B_2(d)\rangle|$ vs. distance $d$, respectively. The black dashed lines correspond to the plot of the exponentially decaying function $\propto\exp(-d/\xi)$ with correlation length $\xi=1.23$ estimated from the gap of the transfer matrix.}
\label{Figure3}
\end{figure}

To identify dominant local correlations in the tRVB state, we have directly calculated several spin-spin and trimer-trimer correlation functions with the fixed-point boundary iMPSs. 
The spin-spin correlation functions include $|\langle \lambda^3(0)\lambda^3(d)\rangle|$ and $|\langle \lambda^8(0)\lambda^8(d)\rangle|$, where $\lambda^3$ and $\lambda^8$ are two diagonal SU(3) Gell-Mann matrices,
\begin{equation*}
\lambda^3=\left(\begin{matrix}1&0&0 \\
0&-1&0 \\
0&0&0 \end{matrix}\right), \quad
\lambda^8=\frac{1}{\sqrt{3}}\left(\begin{matrix}1&0&0 \\
0&1&0 \\
0&0&-2 \end{matrix}\right),
\end{equation*}
and $d$ stands for the distance between the operators in the vertical direction. As shown in Fig.~\ref{Figure3}(a) and (b), these numerically obtained spin-spin correlation functions are equal up to a high precision, which is in agreement with the SU(3) symmetry, and decay with momentum $(\pi, \pi)$ exponentially at large distance, $|\langle \lambda^3(0)\lambda^3(d)\rangle|=|\langle \lambda^8(0)\lambda^8(d)\rangle| \propto\exp(-d/\xi^\text{s})$, where $\xi^\text{s}\sim 0.57$ for $\chi=147$.
The trimer-trimer correlation functions that we have calculated are $C_1(d)=|\langle B_1(0)B_1(d)\rangle-\langle B_1(0)\rangle\langle B_1(d)\rangle|$ and $C_2(d)=|\langle B_1(0)B_2(d)\rangle-\langle B_1(0)\rangle\langle B_2(d)\rangle|$, where $B_1$ and $B_2$ are projectors (i.e., $|\mathrm{trimer}\rangle\langle\mathrm{trimer}|$) onto two different trimers shown in the inset of Fig.~\ref{Figure3}(c) and (d), respectively. These correlation functions also decay exponentially at large distance, and the correlation lengths for $\chi=147$ are $\xi^\text{t}_1 \sim 1.1$ and $\xi^\text{t}_2 \sim 1.2$, respectively. As a comparison, the upper bound of the correlation length obtained from the TM spectrum (with $\chi=147$) is given by $\xi \sim 1.23$ (see dashed lines in Fig.~\ref{Figure3}). Thus, 
the correlation of one $B_1$-type and one $B_2$-type trimers is dominant in the tRVB state.

\textit{Characterizing topological order} --- The fact that the SU(3) tRVB state does not break $C_{4v}$ and translation symmetries and has only short-range correlations indicates that it is a gapped symmetric ground state of a local SU(3) spin Hamiltonian (e.g., the parent Hamiltonian constructed from the PEPS representation). According to a straightforward SU(3) generalization of the celebrated Lieb-Schultz-Mattis-Hastings-Oshikawa (LSMHO) theorem~\cite{lieb1961,hastings2004,oshikawa2000}, the SU(3) spin model on the square lattice with fundamental representation $\mathbf{3}$ on each site cannot have a non-degenerate gapped ground state preserving both SU(3) and translation symmetries. In other words, a gapped ground state with both SU(3) and translational symmetry must be topologically ordered. Based on this LSMHO argument, the SU(3) tRVB state must be a topological spin liquid.

\begin{figure}
\includegraphics[width=0.49\textwidth]{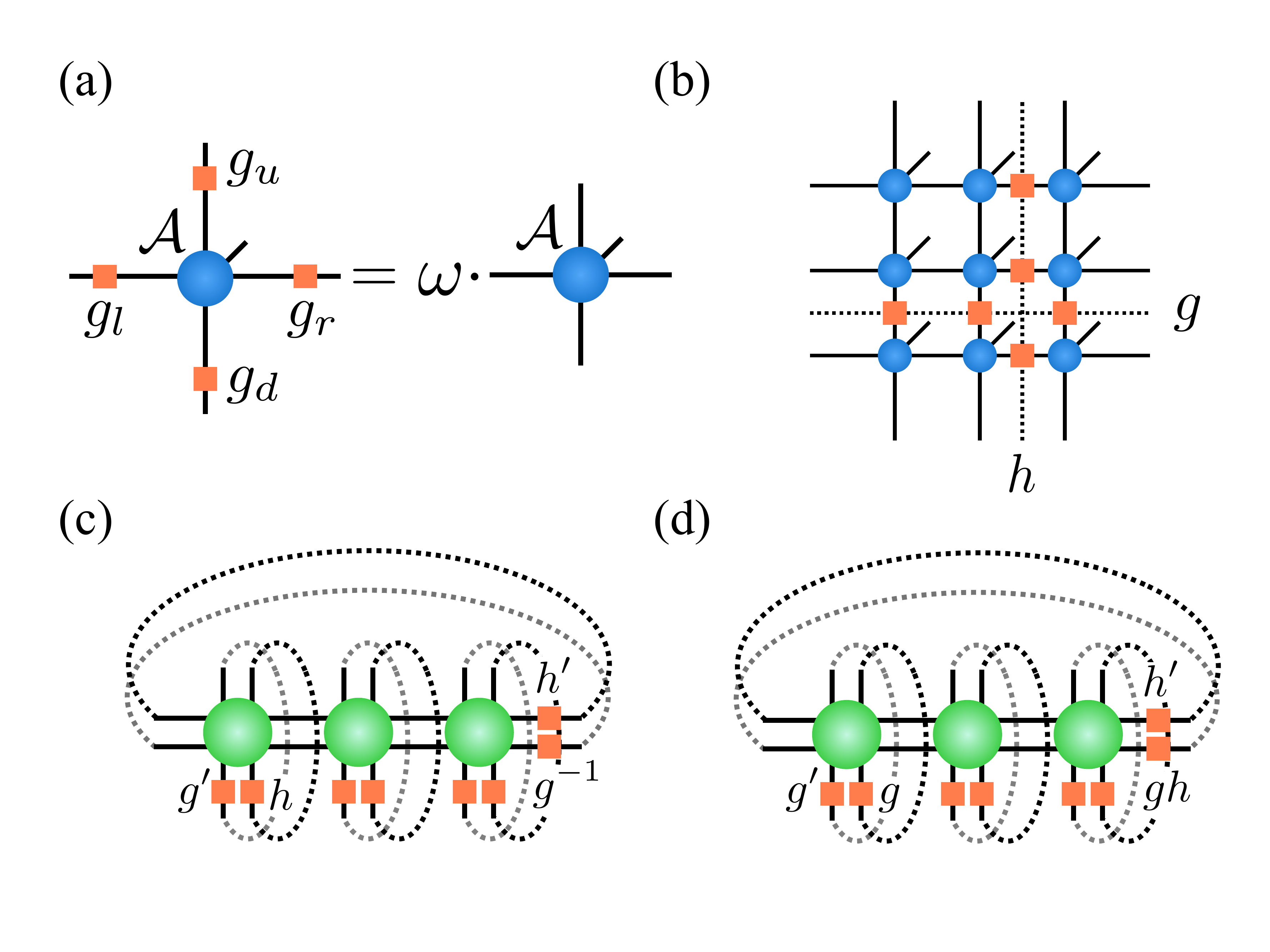}
\caption{(a) $\mathbb{Z}_3$ gauge symmetry of the PEPS local tensor. (b) Constructing nine states by inserting gauge flux along two non-contractible loops on a torus. (c, d) A $3\times 1$ torus formed by double tensors and the $\mathbb{Z}_3$ gauge symmetry elements is used to compute modular $S$ and $T$ matrices.}
\label{Figure4}
\end{figure}

Let us now analyze the topological properties of the tRVB state. An important observation is that the SU(3) tRVB state has a $\mathbb{Z}_3$ gauge symmetry in its PEPS representation. By defining the $\mathbb{Z}_3$-symmetry generator $v=\rm{diag}(1, \omega, \omega, \omega, \omega^2, \omega^2, \omega^2)$ with $\omega=e^{i2\pi/3}$ (the vectors in the virtual Hilbert space are arranged as $|0\rangle,|1\rangle,|2\rangle,|3\rangle,|\bar{1}\rangle,|\bar{2}\rangle,|\bar{3}\rangle$ so that $v$ plays the role of counting the $\mathbb{Z}_3$ charge), the local tensor $\mathcal{A}$ defining the PEPS, given by (\ref{eq:Atensor}), satisfies the following $\mathbb{Z}_3$-injectivity condition~\cite{schuch2010}:
\begin{equation}
(g_l\otimes g_r\otimes g_u\otimes g_d) \mathcal{A} = \omega \mathcal{A},
\label{eq:gaugeSymm}
\end{equation}
where $\mathcal{A}$ is now viewed as a matrix mapping from physical to virtual spaces, and $g_l=g_d=v$ and $g_r=g_u=v^2$, indicating that the left/down/right/up virtual space in each site has a $\mathbb{Z}_3$ gauge symmetry. The symmetry condition (\ref{eq:gaugeSymm}) is graphically shown in Fig.~\ref{Figure4}(a). Notice that, $g_lg_r=g_ug_d=\mathbb{I}$ ($\mathbb{I}$ is an identity matrix in the virtual space $\mathcal{V}_A$), as required from the gauge symmetry condition in PEPS~\cite{schuch2010}.

The $\mathbb{Z}_3$ gauge symmetry, together with the absence of any symmetry breaking order (as revealed from correlation functions), finite correlation length and the LSMHO argument, gives an indication that the tRVB state has $\mathbb{Z}_3$ topological order. Indeed, when the PEPS is defined on a torus, inserting gauge transformations $(g, h)$ on virtual indices in both horizontal and vertical directions [see Fig,~\ref{Figure4}(b)] leads to nine states $|\psi(g, h)\rangle$ in total ($g$ and $h$ can separately take the choice of $\mathbb{I},v,v^2$), which form the nine-fold degeneracy of $\mathbb{Z}_3$ topological order on a torus. However, the linear independence of the nine states is not guaranteed and requires a careful numerical check, which we address below.

To verify the nine-fold ground-state degeneracy and characterize the topological order, we utilize the tensor renormalization group (TRG) method~\cite{levin2007} to compute the modular $S$ and $T$ matrices, which can be viewed as order parameters for topological phases~\cite{he2014}. The TRG process is essentially to compute the overlap of the nine states by first real-space coarse graining the double tensor to a fixed-point tensor, and then contracting a small cluster formed by the fixed-point tensor and gauge transformations [see Fig.~\ref{Figure4}(c, d)].
At each renormalization group (RG) step, a truncation has to be introduced to avoid the exponential growth of bond dimensions, which is achieved by keeping $\chi$ singular values in each RG step.

In order to compute the modular matrices, it is important to keep track of the $\mathbb{Z}_3$ gauge symmetry in each layer, which can be achieved through a simple block singular-value decomposition (SVD) method~\cite{he2014}. After each RG step, we apply the following formula to compute modular $S$ and $T$ matrices:
\begin{equation}
\begin{split}
&\langle \psi(g',h')|\hat{S}|\psi(g,h)\rangle = \langle\psi(g',h')|\psi(h,g^{-1})\rangle,\\
&\langle \psi(g',h')|\hat{T}|\psi(g,h)\rangle = \langle\psi(g',h')|\psi(g,gh)\rangle,
\end{split}
\end{equation}

Compared to the TRG method for the $\mathbb{Z}_3$ deformed toric code state~\cite{huang2015}, there is one subtlety for the tRVB case: the site tensor carries a $\mathbb{Z}_3$ gauge charge 1 or 2, depending on the number of RG steps. This arises from the fact that the physical degree of freedom belongs to the SU(3) fundamental representation and, in each RG step, effectively two sites are merged together. This is different from the spin-1/2 RVB state with $\mathbb{Z}_2$ topological order, where after one RG step, the double tensor becomes charge neutral~\cite{chen2018}. The gauge charge not only influences how we separate the double tensor into different blocks, but also has important consequence in the process of computing the wave-function overlap. In fact, we need to compute wave-function overlaps on the $3\times 1$ or $3\times 3$ cluster where a charge neutral object is formed, instead of a $1\times 1 $ cluster which has gauge charge 1 or 2. Computationally, contracting the bi-layer tensor network on a $3\times 3$ cluster on a torus is more challenging than a $3\times 1$ torus. Thus, we utilize the $3\times 1$ torus for computing the modular matrices [see Fig.~\ref{Figure4}(c, d)].
Another point worth mentioning is that for calculating wave-function overlaps, gauge transformation generator in the upper layer and down layer should be complex conjugate of each other.

\begin{figure}
\includegraphics[width=0.39\textwidth]{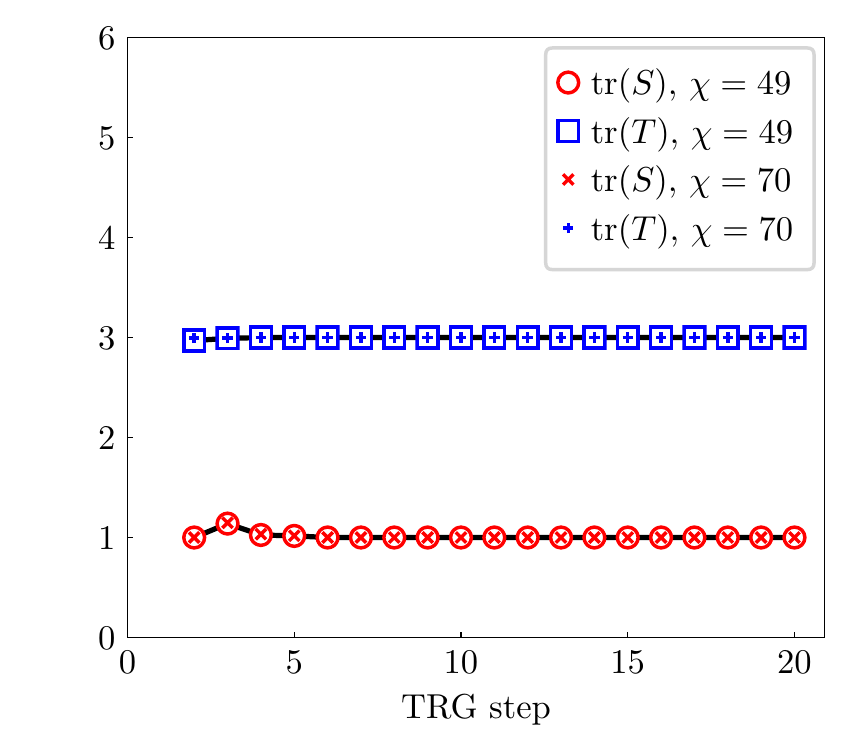}
\caption{Modular matrices from TRG. With increasing RG steps, the modular matrices converge to their fixed point results, as seen from their traces.}
\label{Figure5}
\end{figure}

For the tRVB state, with $\chi=49$ and $70$, within 5 RG steps, the double tensor has flowed to its fixed point, as revealed by the changes in modular matrices during RG (see Fig.~\ref{Figure5}). This quick convergence is in agreement with the short correlation length.
Eventually, the converged modular $S$ and $T$ matrices are found to be
\begin{equation}\label{eq:Z3ST}
\begin{split}
S = \left(\begin{smallmatrix}1&0&0&0&0&0&0&0&0 \\
0&0&0&0&0&0&1&0&0 \\
0&0&0&1&0&0&0&0&0 \\
0&1&0&0&0&0&0&0&0 \\
0&0&0&0&0&0&0&1&0 \\
0&0&0&0&1&0&0&0&0 \\
0&0&1&0&0&0&0&0&0 \\
0&0&0&0&0&0&0&0&1 \\
0&0&0&0&0&1&0&0&0 \end{smallmatrix}\right),
T = \left(\begin{smallmatrix}1&0&0&0&0&0&0&0&0 \\
0&1&0&0&0&0&0&0&0 \\
0&0&1&0&0&0&0&0&0 \\
0&0&0&0&0&1&0&0&0 \\
0&0&0&1&0&0&0&0&0 \\
0&0&0&0&1&0&0&0&0 \\
0&0&0&0&0&0&0&1&0 \\
0&0&0&0&0&0&0&0&1 \\
0&0&0&0&0&0&1&0&0 \end{smallmatrix}\right).\\
\end{split}
\end{equation}
The linear independence of the nine states and the $\mathbb{Z}_3$ topological order are unambiguously demonstrated with above modular $S$ and $T$ matrices.

\textit{Summary and discussion} --- In summary, we have proposed an SU(3) trimer resonating-valence-bond state with $C_{4v}$ point-group symmetry on the square lattice and have characterized it as a $\mathbb{Z}_3$ gapped spin liquid by using the projected entangled-pair state representation. The gap in the transfer matrix demonstrates that all (connected) local correlation functions decay exponentially, and the dominate short-range correlation is found to be one type of the trimer-trimer correlation. The topological order has been established by constructing the topological sectors and calculating the modular $S$ and $T$ matrices, which are in perfect agreement with the $\mathbb{Z}_3$ topological order.

For the SU(2) spin-1/2 nearest-neighbor RVB state, it is known to be a gapless spin liquid on bipartite lattices~\cite{albuquerque2010,tang2011} and a $\mathbb{Z}_2$ topological spin liquid on nonbipartite lattices~\cite{moessner2001,poilblanc2012,schuch2012}, so the interplay between the on-site physical symmetry and the underlying lattice symmetry plays an important role in determining the nature of the state. For the SU(3) tRVB state, further investigations are still needed to characterize it on lattices other than the square lattice.

It is also natural to ask whether there is a realistic Hamiltonian with the SU(3) tRVB state (or its nearby $\mathbb{Z}_3$ spin liquid) being its ground state. The identification of a realistic Hamiltonian stabilizing the $\mathbb{Z}_3$ spin liquid would be quite useful for designing experimental simulation setups with cold atoms in optical lattices. Furthermore, it would also be interesting to study in a microscopic model the competition between the $\mathbb{Z}_3$ spin liquid and other candidate ground states found in SU(3) spin models~\cite{bauer2012,corboz2013,wu2016,pimenov2017}.

\textit{Note added} --- During the preparation of this manuscript, we became aware of an article~\cite{kurecic2018} on the PEPS construction of an SU(3) spin liquid on the kagome lattice.

\textit{Acknowledgment} --- We are grateful to J. Chen, M. Cheng, K. Penc, D. Poilblanc, F. Pollmann, N. Schuch, and Q.-R. Wang for helpful discussions. This work was supported by the TNSTRONG ANR grant of the French Research Council (J.Y.C.) and the DFG via project A06 of SFB 1143 (H.H.T.).

\appendix

\bibliography{Trimer}

\end{document}